\def\daga#1{{#1\mkern -10.0mu /}}
\def\'#1{\if#1i{\accent19\i}\else{\accent19#1}\fi}
\newcommand{\be}{\begin{eqnarray}}
\newcommand{\ee}{\end{eqnarray}}
\documentstyle[preprint,aps]{revtex}
\tightenlines
\begin{document}
\draft
\title{Right handed neutrino production in dense and hot plasmas}
\author{Alejandro Ayala and Juan Carlos D'Olivo}
\address{Instituto de Ciencias Nucleares\\
         Universidad Nacional Aut\'onoma de M\'exico\\
         Apartado Postal 70-543, M\'exico D.F. 04510, M\'exico.}
\author{Manuel Torres}
\address{Instituto de F\'{\i}sica\\
         Universidad Nacional Aut\'onoma de M\'exico\\
         Apartado Postal 20-364, M\'exico D.F. 01000, M\'exico.}
\maketitle
\begin{abstract}

For Dirac neutrinos with magnetic moment, we compute the production rate
for right-handed neutrinos in a hot and dense QED plasma containing an 
initial population of left-handed neutrinos thermally distributed. The most 
important mechanisms for $\nu_L$ depolarization, or production of 
right-handed neutrinos, are the  $\nu_L \to \nu_R$ chirality flip
and the plasmon decay to $\bar{\nu}_L + \nu_R$. The rates for these processes 
are computed in terms of a resummed photon propagator which 
consistently incorporates the background effects to leading order. Applying 
the results to the cases of supernovae core collapse and the primordial 
nucleosynthesis in the early universe, we obtain upper limits on the neutrino 
magnetic moment.

\end{abstract}
\pacs{PACS numbers: 11.10.Wx, 14.60.Lm, 95.30.Cq, 97.60.Bw}

\section{Introduction}\label{intro}

The properties of neutrinos have become the subject of
an increasing research effort over the last years. Among these properties,
the neutrino magnetic moment $\mu_\nu$ has received attention in connection
with various chirality flip processes that could have important consequences
for the explanation of the solar neutrino problem \cite{cisne,volo1},
the dynamics of stellar collapse~\cite{nussi,barbi} and the evolution
of the early universe~\cite{fields}.
A non-vanishing neutrino magnetic moment implies, for example, that 
left-handed neutrinos produced inside a supernova core during the collapse,
could change their chirality becoming sterile with respect to the weak 
interaction. These sterile neutrinos would fly away from the star leaving 
essentially no energy to explain the observed luminosity of the supernova. The 
chirality flip could be caused by the interaction with an external magnetic 
field or by the scattering with charged fermions in the background, for 
instance $\nu_L \, e^{-}  \to \nu_R\, e^{-}$ and $\nu_L \,  p  \to \nu_R\, p$. 
Invoking this last mechanism and using the average parameters inferred from
the supernova 1987A, Barbieri and Mohapatra~\cite{barbi} have derived a 
limit $\mu_\nu < (0.2-0.8)\times 10^{-11} \mu_B$, where $\mu_B$ is the
Bohr magneton.

It has also been pointed out that the constraints imposed by big-bang 
nucleosynthesis (BBN) do not allow the extra degree of freedom that a 
right-handed neutrino in equilibrium would introduce. In order to avoid that 
the chirality flip processes maintain a population of right-handed neutrinos in 
equilibrium during the evolution of the early universe, it is necessary that 
the average rate for these processes is less than the expansion rate of the 
universe at all times until the BBN epoch. Invoking this constraint, Elmfors 
{\it et. al.}~\cite{elmfors} have derived a cosmological bound on the neutrino 
magnetic moment $\mu_\nu < 6.2\times 10^{-11} \mu_B$.

Dispersion processes in a plasma could exhibit infrared divergences due to the 
long-range electromagnetic interactions. To prevent such divergences,
the authors in Ref.~\cite{barbi} introduced an ad hoc thermal mass into the 
vacuum photon propagator. However, it is well known that at high temperatures
or densities, a consistent formalism developed by Braaten and 
Pisarski~\cite{brat1,lebellac} and that renders gauge independent results, 
requires the use of effective propagators and vertices that resum the 
leading-temperature corrections. The method has been successfully applied to 
the study of the damping rates and energy losses of particles propagating 
through hot plasmas~\cite{brat2,brat3,brat4}. In this paper we use this
framework to study the neutrino chirality flip processes in a dense and hot 
plasma.
 
The most efficient process for conversion of left-handed to right-handed 
neutrinos happens through scattering off electrons with the exchange of 
effective space-like photons. In the resummation method of Braaten and 
Pisarski, these photons are described by the spectral function of the photon 
propagator that develops a non-vanishing contribution for space-like momenta 
and whose physical origin is Landau damping. We compute the 
production rate of $\nu_R$'s and the corresponding luminosity for such a 
process in a supernova. Our complete leading-order calculation is compared 
with the results obtained by means of an screening prescription used in 
a previous work~\cite{barbi}. Our result can be used to place an upper bound on
the neutrino magnetic moment which is in the range 
$\mu_\nu < (0.1-0.4)\times 10^{-11}\mu_B$~\cite{nos}.

This work is organized as follows: In section~\ref{Form}, we collect the 
ingredients that allow to compute the production rate of right-handed
neutrinos from the imaginary part of the right-handed neutrino self-energy,
in the real-time formulation of Thermal Field Theory (TFT), by means of a 
resummed photon propagator. In section~\ref{flip}, we restrict the analysis to 
the production of right-handed neutrinos from the chirality-flip process. We
obtain approximate expressions for the production rate that permit us to
explore its analytical behavior in the small and large right-handed neutrino
energy regions. In section~\ref{decay}, we study the production of right-handed 
neutrinos through the plasmon decay process which we show to be subdominant 
as compared to the chirality flip process. In section~\ref{super} we use
the average parameters inferred from the supernova 1987A to find an
upper bound to the neutrino magnetic moment. In section~\ref{early}, we
also deduce an upper bound by imposing that the average production rate of
right-handed neutrinos be at all times less than the Hubble rate up to the
BBN epoch. This last result is shown to differ from that of Ref.~\cite{fields}.
We sumarize our results in section~\ref{conclu} and leave for the
appendices some of the computations outlined throughout the rest of the work. 

\section{Formalism}\label{Form}

Consider a QED plasma in thermal equilibrium at a temperature $T$  such that 
$T\, , \tilde{\mu}_e \gg m_e$, where $m_e$ and $\tilde{\mu}_e$ are the 
electron mass and chemical potential, respectively.
The production rate $\Gamma$ of right-handed neutrinos with total energy 
$E$ and momentum  $\vec{p}$ can be conveniently expressed in terms of the 
$\nu_R$ self-energy $\Sigma$ as~\cite{weldon}
\be
   \Gamma\left(E\right)= \, \frac{n_F \, (E)}{2E} {\mbox T} {\mbox r}
   \, \left[ \daga{P} \, R \,   {\mbox I}{\mbox m}\,  \Sigma
   \, \right]   \label{rate}\, ,
\ee
where $L,R = {1\over 2} \left(1 \pm \gamma_5\right)$ and $n_F$ is the Fermi-Dirac distribution 
for the right-handed neutrino. As shown below, in Eq. (\ref{self1}), the  $\nu_R$ thermal 
distribution cancels out from the final result. 
In what follows we consider $E > 0$ corresponding to $\nu_R$ production.
The annihilation of $\nu_R$ can be obtained from the case $E < 0$, however 
the initial $\nu_R$ population is negligible and the corresponding rate can be
ignored. 

The resummation scheme is usually presented in the imaginary-time formalism. 
However, the expression for Im$\Sigma$ can be directly computed following 
either the imaginary or the real-time formulations of TFT 
with identical results. In what follows we will work in the 
real-time formalism. As is well known, this formalism requires a 
doubling of the degrees of freedom and the propagators and self-energies adopt 
a $2\times 2$ matrix structure. In particular, the imaginary part of the 
retarded self-energy is related to the $1$-$2$ component of the self-energy 
matrix through~\cite{land}
\be
   {\mbox I}{\mbox m}\,  \Sigma (P) \, = \,  
   {\epsilon (E) \over 2i \, n_F \, (E)}  \, 
   \Sigma_{12} (P)
   \label{self1} \, , 
\ee
where $\epsilon(E) = \theta(E) - \theta(-E)$, with $\theta$ the step function. 
As stated above the $\nu_R$ thermal distribution cancels out when Eq. (\ref{self1})
is substituted in Eq. (\ref{rate}).
We take $\sigma = 0$ for the time path-parameter in the notation of Le 
Bellac~\cite{lebellac} and Landsman and van Weert~\cite{land}. 
For simplicity we have selected the rest frame of the medium; however, the 
expressions can be rewritten in a covariant way replacing $E$ by $p \cdot u$, 
where $u_\mu$ is the velocity four-vector of the medium.

\newpage

\let\picnaturalsize=N
\def\picsize{3.0in}
\def\picfilename{figu1.eps}
\ifx\nopictures Y\else{\ifx\epsfloaded Y\else\input epsf \fi
\let\epsfloaded=Y
\centerline{\ifx\picnaturalsize N\epsfxsize \picsize\fi \epsfbox{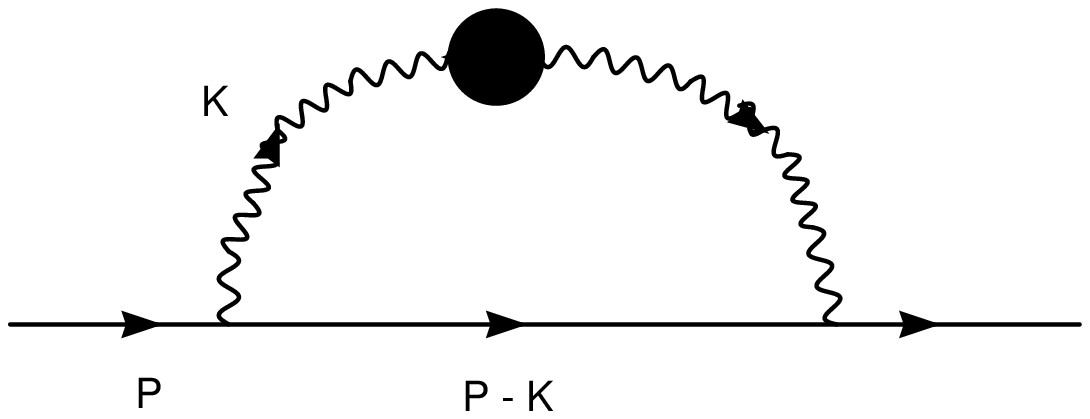}}}\fi
{\it Fig. 1.  Feynman diagram for the  self energy $\Sigma$  of the  right-handed
         neutrino  containing the  effective photon propagator
         (resummed in the HTL approximation)  denoted by a blob.  }
\vskip1.0cm

The one loop contribution to $\Sigma_{12}$, shown in Fig.~1, is given 
explicitly by
\be
   \Sigma_{12}(P) \,=\, -i \mu_\nu^2 \int\frac{d^4K}{(2\pi)^4}
   \,  K_\alpha \,  \sigma^{\alpha\rho}\,  
   S_{12} \left( \daga{P}  \, + \,    \daga{K} 
   \right)  \, L  \,   K_\beta \,  \sigma^{\beta\lambda}
    \,\, ^\star D^{21}_{\rho\lambda}(K)
   \label{self2} \, .
\ee

We will use capital letters to denote four-vectors: $P^\mu  = ( E, \vec p)$
for the incoming neutrino and $K^\mu  = ( k_0, \vec k)$ for the virtual photon,
and $p  \equiv |\vec p|$, $k \equiv |\vec k|$. For the neutrino-photon vertex
we take the magnetic dipole interaction $\mu_\nu  \sigma_{\alpha \beta}
K^{\beta}$. The neutrino effective electromagnetic vertices are of no concern
to us here since they are induced by the weak interaction of the
particles in the background and thus conserve chirality~\cite{JCNP}.

The intermediate $\nu_L$ line can  be taken as a bare fermion propagator 
because it gets dressed only through weak interactions with
the particles in the medium. Hence, the $S_{12}$ component for a massless 
fermion propagator is  given by 
\be
   S_{12}(Q) \, = \, 2 \pi i  \, \daga{Q}\, \delta \left( Q^2 \right)
   \epsilon(q_0) \,  n_F(q_0)
   \label{propf} \, ,
\ee
where
\be
   n_F(q_0)  \, = \, {1 \over e^{(q_0 - \tilde{\mu}_\nu )/T} + 1}\, 
\ee
is the  Fermi-Dirac distribution for the left-handed neutrino in the 
medium, with $\tilde{\mu}_\nu$ being its chemical potential.

In Eq. (\ref{self2}), the integration region where the momentum
$k$ flowing through the photon line is soft ($i.e.$ of order $eT$) requires 
hard thermal loop (HTL) corrections to the photon propagator that contribute 
at leading order and must be  resummed. The effective propagator is 
represented by the blob in Fig.~1, and is obtained by summing the geometric 
series of one-loop self-energy  corrections proportional to $e^2 T^2$.   
As usual, we split the photon propagator into longitudinal and transverse 
parts 
\be
   ^\star D^{\mu\nu}(K) &=&  ^\star D_L(K) \, {P^{\mu\nu}_L} \, + \,
   ^\star D_T(K) \, {P^{\mu\nu}_T}   \label{prop} \, ,
\ee
we drop the term proportional to the gauge parameter since
it does not contribute to $\Sigma$, as can be easily checked. In the previous 
equation all $^\star D^{\mu\nu}$,  $^\star D_T$ and $^\star D_L$ are 
$2\times 2$ matrices, while $P^{\mu\nu}_L$ and $P^{\mu\nu}_T$ are the  
longitudinal and transverse projectors, respectively, 
\be 
 P^{00}_T &=&   P^{0i}_T = 0 \, , \qquad  P^{ij}_T = \delta^{ij} - \hat{k}^i \hat{k}^j \, , 
\nonumber \\
 P^{\mu\nu }_L &=& - g^{\mu\nu} + {K^\mu K^\nu  \over K^2 }
-  P^{\mu\nu }_T  
\label{projectors}\, .
\ee

The effective photon propagator is obtained when the hard thermal loops in the 
photon self energy (first computed by Klimov~\cite{klim} and 
Weldon~\cite{weldon2}) are resummed. The complete matrix propagator is written 
as 
\be
   ^\star D_{L,T}  (k) = \tilde{U} \pmatrix{
   {1 \over k^2 - \Pi_{L,T} + i \epsilon}  & 0 \cr
   0 &  {-1 \over k^2 - \Pi_{L,T}^* - i \epsilon} \cr
   } \tilde{U} 
   \label{prop1}   \,  , 
\ee
where $\tilde{U}$ is the photon thermal matrix~\cite{land}. The polarization 
functions $\Pi_L$ and $\Pi_T$  are given by 
\be
   \Pi_L  (K) &=& - {2 m^2_\gamma K^2 \over k^2 } \left[
   1 - {k_0 \over k} Q_0 \left({k_0 \over k} \right) \right]\, ,\nonumber \\ 
   \Pi_T  (K) &=&   {m^2_\gamma  k_0 \over k } \left[ {k_0 \over k} + 
   \left( 1 - \left({k_0 \over k} \right)^2 \right) Q_0  
   \left({k_0 \over k}  \right)   
   \right] 
   \label{prop2}   \,  , 
\ee
$m_\gamma$ is the photon thermal mass, that in the limit $T,\tilde{\mu}_e \gg 
m_e$ is given by
\be
   m_\gamma^2=\frac{e^2}{ 6 }\left(  T^2 + {3  \tilde{\mu}_e^2  \over \pi^2}
   \right)
   \label{mass} \, .
\ee
The Legendre function $Q_0(k_0/k)$ is defined in the complex $k_0$ plane cut 
from $-k$ to $k$; it is real in the time-like region but it acquires an 
imaginary part for space-like $K$ 
\be
   Q_0\left({k_0 \over k}\right)=\frac{1}{2} 
   \ln\left|\frac{k_0 + k}{k_0 - k}\right|
   - i \frac{\pi}{2}\, \theta(k^2 - k_0^2)
   \label{legen}   \,  . 
\ee
The solution of the equations $K^2 - Re \Pi_L= 0 $  and $K^2 - Re \Pi_T= 0 $
represent the propagation of longitudinal photons, or plasmons, and transverse
photons, respectively. These solutions in the time-like region are well
known,   in general they are obtained numerically. 
However it is possible to obtain approximate analytical results for small and 
large values of $k$.  In the small momentum limit $k \ll m_\gamma $ the 
dispersion relations reduce to 
\be 
   \omega^2_T \, &=& \, \omega_p^2 + {6 \over 5} k^2
   \, , \nonumber \\
   \omega^2_L \, &=& \, \omega_p^2 + {3 \over 5} k^2
   \, ,  \label{reld1}
\ee
where the plasma frequency is defined as 
$\omega_p = \sqrt{ {2\over 3} }m_\gamma$. In the large  momentum limit 
$k \gg  m_\gamma $ the  behavior of the dispersion relations is 
approximated by 
\be 
   \omega^2_T \, &=& \,  m_\gamma^2 + k^2 
   \, , \nonumber \\
   \omega_L \, &=& \,  k +  2 k  \exp{\left( - {k^2 \over
   m_\gamma^2}
   \right)}
   \, .  \label{reld2}
\ee

In a relativistic plasma  the photon dispersion relation for the longitudinal 
mode $K^2 - Re \Pi_L= 0 $ has also a solution in the space-like 
region~\cite {jc}. Then the \v{C}erenkov radiation of a plasmon is, in 
principle, kinematically  allowed. The numerical solution of the longitudinal 
dispersion for $\omega  < k$ is shown in Fig.~2. We observe that the solution 
is close to the light cone. Using this fact and the first of 
Eqs.~(\ref{prop2}), it is a simple task to derive the following approximate 
solution 
\be 
\omega_L  =  k - 2 k {1 \over  1 +  \exp{\left(  {k^2 + 2  m_\gamma^2 \over
   m_\gamma^2}  \right) } } 
\label{reldsl} \, , 
\ee
that, as shown in  Fig.~2, agrees very well with the numerical solution.
However, this mode develops a large imaginary part, which implies that the 
Landau damping  mechanism acts to preclude its propagation. In this situation 
the correct method to include  the complete contribution of both the 
space-like and time-like degrees of freedom requires the use of the spectral 
representation as given below in Eq.~(\ref{rolong}). 
The effect of the space-like mode is not very distinctive, except for the low 
energy  spectrum  of the $\nu_R$ production via the spin flip reaction 
$\nu_L \to \nu_R$ (see next section).

\vskip1.0cm

\let\picnaturalsize=N
\def\picsize{5.0in}
\def\picfilename{figu2.eps}
\ifx\nopictures Y\else{\ifx\epsfloaded Y\else\input epsf \fi
\let\epsfloaded=Y
\centerline{\ifx\picnaturalsize N\epsfxsize \picsize\fi \epsfbox{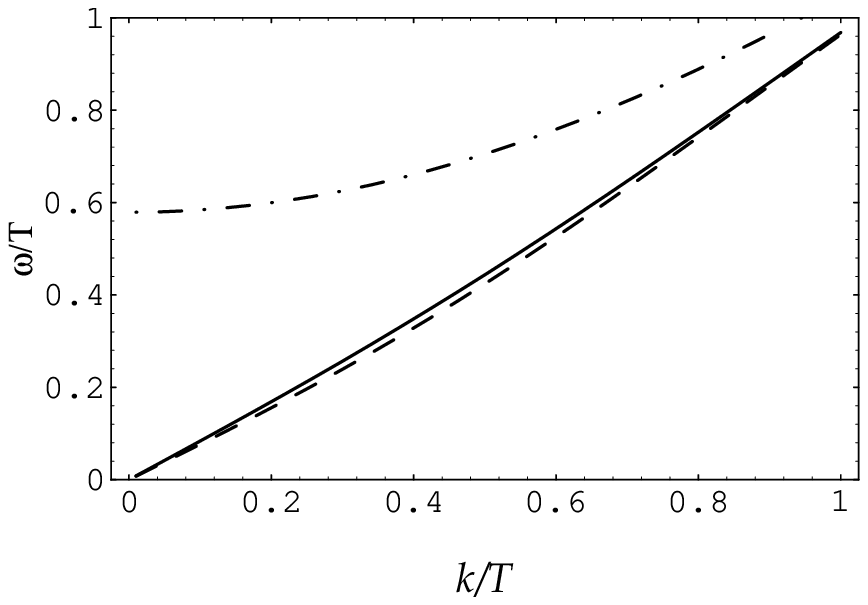}}}\fi

{\it Fig. 2.  Dispersion relations for the longitudinal modes. The upper branch 
         (dotted -dashed line) is the the usual time-like solution. The 
         lower brach is the space-like mode. The numerical solution 
         (solid line) is compared with the approximate analytical result 
         (dashed line). }
\vskip1.0cm

From Eqs. (\ref{prop1}) and (\ref{prop2}) we can get the component 
$^\star D^{21}$of the photon propagator. In order to obtain the complete 
result, we notice that for time-like momenta the $\Pi_L$ and $\Pi_T$ functions
are real, however for space-like momenta both functions acquire an imaginary 
part as seen from Eqs.~\ref{prop2} and~\ref{legen}. The final result can be 
written as 
\be 
   ^\star D_{L,T}^{21} =  2\pi i \left[ 1 + f(k_0)\right] \rho_{L,T}(k_0\,,k)
   \, ,  \label{fotopro}
\ee
where $f(k_0)=(e^{k_0/T}-1)^{-1}$ is the Bose-Einstein distribution and the 
functions  $\rho_{L,T}(k_0\,,k)$ are given by 
\be
   \rho_i(k_0\,,k)=  \, 
   Z_i\left( k\right)\left[  \delta 
   \left( k_0 - \omega_i(k) \right) + \delta 
   \left( k_0 + \omega_i(k)\right)   \right]
   +  \beta_i(k_0,k) \theta \left(k_0^2 - k^2 \right)
   \, ,
   \label{rolong}
\ee
for $i=L,T$. The residue $Z_L(k)$ for the longitudinal excitations is given by 
\be
   Z_L(k)  =  \,   
   {\omega_L(k) \over k^2 + 2 m_\gamma^2 -  \omega_L^2(k)}
   \, ,
   \label{resilong}
\ee
and the cut function $\beta_L(k_0\,,k)$
\be
   \beta_L(k_0\,,k)=  \, \left(\frac{x}{x^2 - 1}\right) \,
   \frac{m_\gamma^2}
   {\left[k^2+2m_\gamma^2 \, (1-\frac{x}{2}\ln\left|\frac{x+1}{x-1}\right|)
   \right]^2+\left[\pi m_\gamma^2\,  x\right]^2}\, ,
   \label{bet1}
\ee
where we have defined $x\equiv  k_0/k$.  For the transverse  mode,
the residue and cut functions are given by
\be
   Z_T(k)  =  \,   
   {\omega_T(k) \left( \omega_T^2(k) - k^2 \right) \over
   2 m_\gamma^2 \omega_T^2(k) - \left( \omega_T^2(k) - k^2 \right)^2 }
   \, ,
   \label{resitran}
\ee
\be
   \beta_T(k_0\,,k)= \, \frac{(1/2)  
   m_\gamma^2 \, x \, \left(1 - x^2 \right)}
   {\left[ k^2(1 -x^2)+m_\gamma^2 \, \left( x^2+
   \frac{x}{2}(1 - x^2)\ln\left|
   \frac{x+1 }{x-1}\right|\right)\right]^2 +
   \left[\frac{\pi}{2} m_\gamma^2\,  x \left(1 - x^2\right)\right]^2} \, .
   \label{bet2}
\ee
The functions $\rho _{L,T}(\omega\,,k)$ are referred to as the photon spectral
densities. It is an straightforward exercise to verify that these coincide with
the spectral densities computed in the imaginary-time formalism after the 
analytical continuation to real time is performed. The spectral densities 
$\rho_{L,T}(\omega\,,k)$ contain the discontinuities of the photon propagator 
across the real-$\omega$ axis. Their support depends on the magnitude
of the ratio between $\omega$ and $k$. For $\left| \omega/k \right| > 1$,
$\rho_{L,T}(\omega\,,k)$ have support on the points $\pm \omega_{L,T}(k)$,
i.e., the time-like quasiparticle poles. In the space-like region the support 
of $\rho_{L,T}(\omega\,k)$ lies on  the whole interval $-k<\omega<k$, with the
contribution arising from the branch cut of $Q_0$.  Hence, the  spectral 
density is the sum of pole and cut terms.

To proceed with the calculation of Eqs.~(\ref{rate}) and (\ref{self2}) we need 
the following quantities
\be
   C_T(E,K) \equiv K_{\alpha}K_{\beta}P_{T\mu\nu}\,{\mbox T}{\mbox r}
   \left[\sigma^{\alpha\mu}( \daga{P} + {K \mkern -13.0mu / }
   ) \, L\, \sigma^{\beta\nu} \daga{P}
   \, R\,  \right]&=&
   k^2 \left( x^2 - 1\right)^2 \left[(2E+\omega)^2-k^2\right]\, ,
   \nonumber\\
   C_L(E,K) \equiv   K_{\alpha}K_{\beta}P_{L\mu\nu}\,{\mbox T}{\mbox r}
   \left[\sigma^{\alpha\mu}( \daga{P}  \,  + {K \mkern -13.0mu / } )
   \, L \, \sigma^{\beta\nu} \, \daga{P}      \, R \,  \right]
   &=&  - k^2 \left( x^2 -1 \right)^2 \left(2E+\omega \right)^2\,  .
   \label{trace}
\ee
The production rate of right-handed neutrinos in Eq.~(\ref{rate}) can be
computed after the substitution of the expressions for the fermion, 
Eq.~(\ref{self1}), and photon, Eq.~(\ref{fotopro}), propagators into 
Eq.~(\ref{self2}). Using also the spectral representations, Eq.~(\ref{rolong}),
our result is 
\be
   \Gamma(E) &=& \frac{\mu_\nu^2 \pi^2}{  E}\int \frac{d^4k}{(2\pi)^4} \,
   \epsilon(E  +  k_0) \epsilon(k_0) \left(1 + f(k_0)  \right) \,
   n_F(E +  k_0 )\nonumber\\
   &&  
   \left[C_L(E,K) \,  \rho_L(k_0,k)+
   C_T(E,K) \, \rho_T(k_0,k)\right]\, .
   \label{game1}
\ee
As we have already mentioned we shall consider $E> 0$ corresponding to the 
production of $\nu_R$. However, depending  on the signs of $E + \omega$ and 
$\omega$  the  $\nu_R$ production rate can be divided into the chirality flip 
process $\nu_L\to\nu_R$ mediated by a virtual photon and the plasmon decay 
process $\gamma \to \bar{\nu}_L \nu_R$. We shall analyze these two 
processes separately in the following sections. 

\section{The $\nu_L \to \nu_R$ chirality  flip }\label{flip}

The contribution to the  $\nu_L \to \nu_R$ chirality flip process is obtained 
from the general expression in Eq.~(\ref{game1}) if we set  $E + \omega >0$ as 
corresponds to the case of an incident $\nu_L$. The angular integration over 
the direction of $\vec k$ is  readily performed. Using the condition 
$|\cos\Theta|\leq 1$, where $\Theta$ is the angle between the momenta of the 
incoming neutrino and the virtual photon, we obtain two kinematical 
restrictions, namely, $|\omega| \leq  k$ and $(k -\omega )\leq 2E$. The first 
condition implies that for the chirality flip $\nu_L \to \nu_R$ process the 
kinematically allowed region is restricted to $|\omega| < k$ and  
consequently the only contribution from the photon spectral density in 
Eq. (\ref{rolong}) arises from the cuts. The rate of production  of 
right-handed neutrinos from  the $\nu_L \to \nu_R$ flip can be computed after  
the substitution of the  expressions for the fermion, Eq.~(\ref{self1}),  and 
photon, Eq.~(\ref{fotopro}), propagators into Eq.~(\ref{self2}), the result is 
\be 
   \Gamma(E) &=& \frac{\mu_\nu^2}{16\pi  E^2}\int_0^{\infty}k  dk \,
   \int_{-k}^{k}d\omega
   \,   \theta(2E+\omega-k)   \left(1 + f(\omega)  \right) \,  n_F(E + \omega
   )\nonumber\\
   && \left[C_L(E,K) \,  \beta_L(\omega,k)+
   C_T(E,K) \, \beta_T(\omega,k)\right]\, ,
   \label{game2}
\ee
where we set $k_0=\omega$. Both, longitudinal and transverse photons, 
contribute to this rate. Notice that using the restriction 
$(k -\omega )\leq 2E$, the integrand in the previous equation can be proved to 
be positive definite. The rate $\Gamma$ can be written as the sum of two 
contributions $\Gamma = \Gamma_e + \Gamma_a$, that correspond to the production
of $\nu_R$ through the emission or absorption of a virtual photon.
$\Gamma_e$  comes from the interval $0\leq \omega < k$, whereas $\Gamma_a$
corresponds to the interval $-k < \omega \leq 0$, as can be checked by means of 
the identity $1+f(\omega)+f(-\omega)=0 $ and the substitution 
$\omega\rightarrow -\omega$  in this second interval.

The results for the longitudinal and transverse contributions to $\Gamma(E)$ as
a function of $E$ are shown in Fig.~3, for a selection of values
characteristic of a supernova core (see section~\ref{super}). The solution for 
the chirality flip production rate requires in general numerical integration.  
However, the physical interpretation of the results is made clearer with 
approximate analytical solutions that  can be derived both in the small and 
large energy limits.

Let us first consider the region of small neutrino energy
$(E \ll T, \tilde{\mu}_\nu)$.  From Fig.~3, we observe that the longitudinal
contribution shows a pronounced peak at energies below $5$ MeV. The result is
somewhat unexpected because in the study of similar processes, the effect of
infrared logarithmic singularities shows up in the transverse contributions.
For example, in the calculation of the lifetime of a fermion quasiparticles in
a $QED$ plasma at high temperature the soft photon contribution leads to an
infrared divergence in the transverse component alone due to the vanishing of
the magnetic mass. The problem was solved with the use of a non-perturbative
method based on a generalization of the Bloch-Nordsieck
approximation~\cite{blaizot}. In the present case, there are enough powers of
$k$ coming from the vertex factors in Eq.~(\ref{self2}) to render the
transverse contribution finite in the infrared. We will show this explicitly in
appendix A. On the other hand the longitudinal photon is known to be screened 
by the Debye mass in the $\omega =0, k \to 0$ limit, yet our results lead to a 
pronounced peak at small $E$. As we shall see, this effect
arises because the longitudinal spectral density, Eq.~(\ref{bet1}), has a
logarithmic divergence at the light cone. A qualitative way to understand
this behavior stems from the observation that there also exists a space-like
branch for the longitudinal mode~\cite{jc}, though with a large imaginary part
that precludes propagation of the mode. This branch lies very close to the
light-cone. Could we ignore the large damping that the mode experiences, the
situation would correspond to a process in which a left-handed neutrino changes
its chirality becoming a right-handed neutrino through the emission or
absorption of a \v{C}erenkov photon. As we cannot ignore damping, the
contribution shows up as that from a resonance that peaks for small
right-handed neutrino energies, that is, for the region where the difference
$k-\omega
{\ \lower-1.2pt\vbox{\hbox{\rlap{$>$}\lower5pt\vbox{\hbox{$\sim$}}}}\ } 0$
is small.

\vskip1.0cm

\let\picnaturalsize=N
\def\picsize{5.0in}
\def\picfilename{figu3.eps}
\ifx\nopictures Y\else{\ifx\epsfloaded Y\else\input epsf \fi
\let\epsfloaded=Y
\centerline{\ifx\picnaturalsize N\epsfxsize \picsize\fi \epsfbox{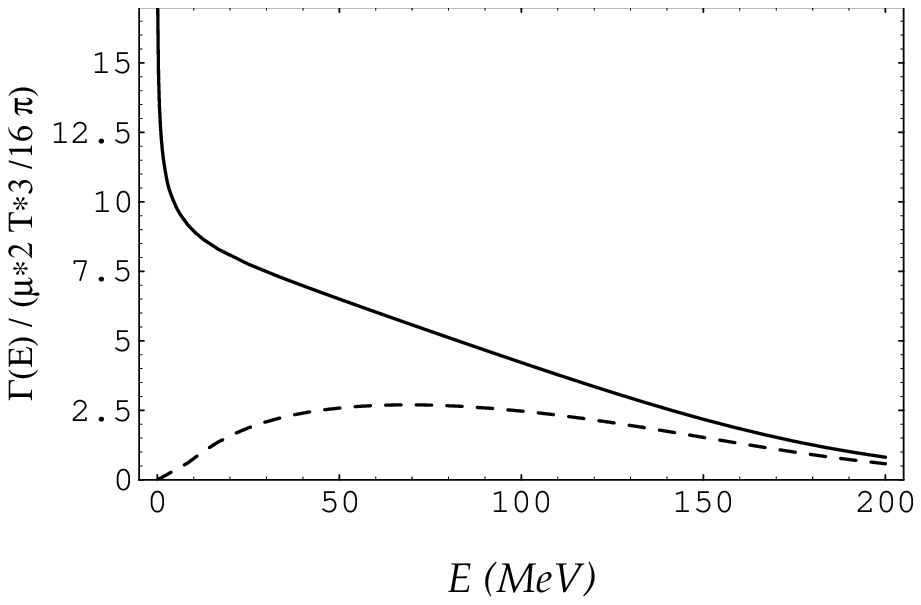}}}\fi

{\it  Fig. 3. The  Longitudinal  (solid  line)  and transverse  contributions  
        (dashed  line) to  the production rate $\Gamma(E)$ for right-handed 
        neutrinos via the  spin flip transition $\nu_L \to \nu_R$. }

\vskip1.0cm

Let us consider the limit $E \to 0$, then the virtual photon is forced to be 
near the light cone 
$(\omega{\ \lower-1.2pt\vbox{\hbox{\rlap{$<$}\lower5pt\vbox{\hbox{$\sim$}}}}\ }
k) $. The integration in Eq.~(\ref{game2}) is restricted to the kinematical 
region $\omega\leq k\leq 2E+\omega$, hence in the limit $E \to 0$  we may set 
$k = E + \omega$ everywhere in the $k-$integration. Keeping only the leading 
term in $E$, the cut contributions to the spectral densities can be replaced by
the approximate expressions
\be 
   \beta_T &=& {E \over \omega \,  m^2_\gamma } \, , \nonumber \\
   \beta_L &=& - {\omega  \over 2 E \,  m_\gamma^2   \,\,  
   \left[ \ln\left(E/T \right)\right]^2 }  \, , 
\ee 
where in the last expression we have simplified the approximation by replacing
$\omega$ by its mean value $T$. We notice that the dependence on the 
longitudinal spectral function that originates the strong rise at small $E$ 
comes from the fact that near the the light cone $\beta_L$ behaves as 
$1/\left[ \ln\left(k - \omega \right) \right]^2$. Using the previous results 
and the identity 
\be 
   \left[1 + f(\omega) \right] [ n_F(E + \omega )] =  n_F(E )
   \left[ f(\omega)  +  n_F(E + \omega)\right]
   \, ,
   \label{equa}
\ee
the resulting integrals  can be evaluated analytically. Separating $\Gamma(E)$ 
into its  transverse and longitudinal parts,  we obtain for the transverse 
contribution  in the limit $E \to 0$
\be 
   \Gamma_T(E)  \approx \frac{\mu_\nu^2 \, E^3  \, T^2 }{\pi  m_\gamma^2 } \, 
   n_F(0) \, \left[ Li_2\left(- e^{\tilde{\mu}_\nu/T}\right) - 
   {\pi^2 \over 6} \right]\, ,
   \label{gap1}
\ee
where $Li_n(z) = PolyLog[n,z]$ is the PolyLog function~\cite{poly}. For the 
longitudinal part the result reads 
\be 
   \Gamma_L(E)  \approx \frac{6 \mu_\nu^2 \,   T^5 }{\pi  m_\gamma^2 } \, 
   { n_F(0) \over \left[ \ln{(E/T)}\right]^2} \, 
   \left[Li_5\left(- e^{\tilde{\mu}_\nu/T}\right) - \zeta(5)\right] 
   \, ,
   \label{gap2}
\ee
where  $\zeta(z)$ is the Riemann zeta function.  Both contributions vanish at 
$E = 0$, however the longitudinal part shows a steep rise for small $E$ that
originates the peak observed in Fig.~3. The maximum for the longitudinal 
spectrum can also be estimated and it is attained for a very small value of 
the energy 
\be 
   E   \, \approx  \,  T \,  \exp 
   \left(-{T^2 \over m_\gamma^2} \right)
   \, .
   \label{gap3}
\ee
The previous results show that  the $\nu_L \to \nu_R$ chirality flip reaction 
produces a huge number of small energy right-handed neutrinos. 
While this effect looks remarkable, the phenomena that it can give rise to are 
probably unobservable because at very low neutrino energies, experimental 
discrimination between the charged current interactions of low energy $\nu_R$ 
and $\nu_L$ would be fairly difficult. We also notice that in this limit, the 
reaction rate $\Gamma(E)$ depends nonlinearly on the coupling constant $e$.

We now turn our attention to the limit in which  the neutrino energies are 
larger or of order $T$, $\tilde{\mu}_\nu$. As we shall see, a very good 
analytical approximation can be obtained in this case. The method extracts 
explicitly the leading logarithmic screening  terms that arise from the use of 
the full resummed propagator.  Following Braaten and Yuan \cite{brat4} we 
introduce and intermediate cut-off $q^*$ such that $m_\gamma \ll  q^* \ll T, 
\tilde{\mu}_\nu$. In the region of hard momentum transfer $k > q^*$ the 
tree-level approximation is used for the virtual photon, $q^*$ acting as an 
infrared regulator. In the soft region $k < q^*$, the effective resummed 
propagator is used. Adding the hard and soft contributions, the dependence on 
the arbitrary scale $q^*$ cancels.

Let us first outline the calculation of the  soft-momentum transfer 
contribution to $\Gamma(E)$. In this region, hard thermal loop corrections to 
the photon propagator are not suppressed by powers of $e$, hence the  resummed 
photon propagator must be used. Restricting our attention to the momentum 
region $k < q^*$  we can set $1 + f(\omega) \approx T/ \omega$ and for the 
functions in Eq.~(\ref{trace}) we make the approximations 
\be 
   C_T(E,K)  \approx  -C_L(E,K)  \approx  {4  K^4 E^2 \over k^2 }
   \,
   \label{traza2}
\ee
inserting the results into  Eq.~(\ref{game2}), this last reduces to 
\be 
   \Gamma_{soft}(E) = \frac{\mu_\nu^2 \, T \,  n_F(E )}{4 \pi  } 
   \int_0^{q^*}k^3  dk \,
   \int_{-k}^{k} {d\omega \over \omega} 
   \,     \left(1 - {w^2 \over k^2}   \right)^2  \, 
   \left[   \beta_T(\omega,k) -  \beta_L(\omega,k) \right]
   \, . 
   \label{game3}
\ee
The integral over $\omega$ can be  evaluated by using the sum rules derived 
from the analytical properties of the effective propagators~\cite{pisars}
\be 
   \int_{-k}^{k}d\omega \omega^{2n - 1} 
   \left(1 - {\omega^2 \over k^2}  \right) \beta_L&=&
   -2 \left(1 - {\omega^2_L \over k^2}  \right) \omega^{2n - 1}_L Z_L + 
   k^{2 n - 2} \left[ {2 m_\gamma^2 \over 2 m_\gamma^2  + k^2}   , 
   {2 m_\gamma^2 \over 3 k^2}  \right]
   \nonumber\\
   && {\rm for} \quad n = 0, 1\, ,
   \label{sum1}
\ee
\be 
   \int_{-k}^{k}d\omega \omega^{2n - 1}  \beta_T&=&
   -2  \omega^{2n - 1}_T Z_T  + 
   k^{2 n - 2} \left[ 1, 1, {2 m_\gamma^2  + 3 k^2 \over  3  k^2}     \right]
   \nonumber\\
   && {\rm for} \quad  n = 0, 1, 2\, .
   \label{sum2}
\ee
The logarithmic dependence on $q^*$ can be extracted analytically, the result 
is 
\be 
   \Gamma_{soft}(E) = \frac{\mu_\nu^2 \, m_\gamma^2  T \, }{2 \pi  }  n_F(E )
   \left( \ln{\left[ {q^* \over \sqrt{2} m_\gamma } \right]}  + 
   \int_0^{\infty} dk \,
   \left[{ (\omega_L^2 - k^2)^2 \over  m_\gamma^2 k \omega_L}  Z_L 
   -  { (\omega_T^2 - k^2)^2 \over  m_\gamma^2 k \omega_T}  Z_T 
   \right] \right)\, ,
   \label{game4}
\ee
where we used the fact that $m_\gamma \ll q^*$ and have extended the limit
of integration to $\infty$ since we are just interested in the leading
contribution. The remaining integral inside the brackets is a dimensionless
quantity and can be computed numerically with the result  $-0.14$. 

We now turn to the  hard-momentum transfer contribution to $\Gamma(E)$. In this
region  the tree-level  photon propagator  must be used. Following  
Le Bellac~\cite{lebellac} we notice that  tree-level results can be recovered  
from the full calculation by simply neglecting $m_\gamma^2$ in the denominators
of Eqs. (\ref{bet1}) and (\ref{bet2}). Hence, the approximated spectral 
densities yield
\be
   - \beta_L \approx 2\beta_T \approx   {m_\gamma^2 \omega /k^3  \over  
   k^2  -  \omega^2    } 
   \label{aprox} \, .
\ee
It is convenient to further  decompose the hard region $(k > q^* )$ into  
a  (I)  low $(\omega <  q^* )$  and  (II) high $(\omega > q^* )$ frequency 
regions.  In the low  frequency region we can still use the approximations 
$1 + f(\omega) \approx T/ \omega$  and Eq.~(\ref{traza2}).
The remaining integrations are readily performed.  Recalling that $q^* \ll E$,
the result is 

\be
   \Gamma^{I}_{hard}(E) = 
   \frac{2 \mu_\nu^2 \, m_\gamma^2  T \, }{3 \pi  }  n_F(E )
   \, . 
   \label{game5}
\ee
In the  high frequency region it is not obvious that we can neglect $\omega$ 
and $k$ as compared to $E$, in particular the approximation in 
Eq.~(\ref{traza2}) may not be accurate. However for 
$E \sim T, {\tilde \mu}_\nu$ the integrals are dominated by  small $\omega$ 
and $k$, hence neglecting $\omega, k$  as compared with  $E$ yields the the 
leading order contribution in  $T/E$. The approximation can be improved  
for large $E$, by computing the next order term, the second  order corrections 
are given in appendix B. Thus to leading order
we can set $\omega , k \to 0$ as compared with $E$ wherever possible, except 
for the Bose and Fermi distributions that provide  the cutoff for the integrals
and must be taken exact. Using the identity in Eq.~(\ref{equa}) we obtain 
\be
   \Gamma^{II}_{hard}(E) &=& \frac{\mu_\nu^2 \, m_\gamma^2  T  }{2 \pi  }  
   n_F(E )\left[  \ln\left( { T\over q^*}  \right) + \ln 
   \left(  1 +  e^{(E - \tilde{\mu}_\nu)/T } \right) 
   + {1 \over 2} 
   \ln \left( { 1 -  e^{-E/T }  \over 
   1 +  e^{- \tilde{\mu}_\nu/T } }\right)\right.\nonumber\\
   &+& \left. {E-\tilde{\mu} \over T}
   \right]\, . 
   \label{game6}
\ee
The complete rate for the production of right handed neutrinos to leading order
in $e$ is the sum of  Eqs.~(\ref{game3}), (\ref{game4}) and (\ref{game5}). The 
result is 
\be
   \Gamma(E) &=& \frac{\mu_\nu^2 \, m_\gamma^2  T  }{2 \pi  }  n_F(E )
   \left[  C +  \ln\left( { T\over \sqrt{2} m_\gamma }  \right)  +
   {3(E - \tilde{\mu}_\nu ) \over 4 T}\right.\nonumber\\  
   &+& \left. {1 \over 2}
   \ln \left( {  \cosh^2 \left( {E - \tilde{\mu}_\nu  \over 2 T} \right) 
   \sinh\left({E \over 2 T} \right)\over 
   \cosh \left({ \tilde{\mu}_\nu  \over 2 T} \right) }\right)
   \right]
   \, ,
   \label{game7}
\ee
where $C = 1.88$. Note that the dependence on the arbitrary intermediate scale 
$q^*$ cancels out between the soft contribution in Eq.~(\ref{game3}) and the 
hard contribution in Eq.~(\ref{game5}). In Fig.~4, we display the comparison of 
this analytical approximation with the result of the numerical integration of  
the exact result in Eq.~(\ref{game2}). We notice that for energies in the 
region $E \gg T , \tilde{\mu}_\nu $ the analytical result gives an excellent 
approximation to the exact result. As mentioned before the result can be 
improved by computing the next order correction in $T/E$. Fig.~4 also shows the
result of the second order contribution calculated in appendix B.

\vskip1.0cm

\let\picnaturalsize=N
\def\picsize{5.0in}
\def\picfilename{figu4.eps}
\ifx\nopictures Y\else{\ifx\epsfloaded Y\else\input epsf \fi
\let\epsfloaded=Y
\centerline{\ifx\picnaturalsize N\epsfxsize \picsize\fi \epsfbox{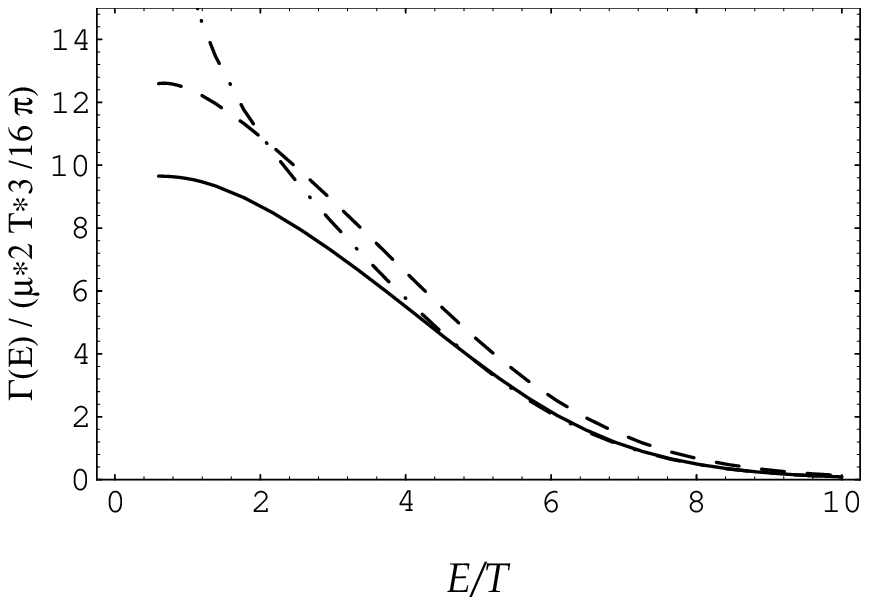}}}\fi

{\it Fig. 4.  The production rate $\Gamma(E)$ for right-handed neutrinos via the
         spin flip transition $\nu_L \to \nu_R$. The exact (numerical) 
         result in the solid line is compared with the approximate result 
         (dashed line) to leading order in $T/E$ and the approximate result 
         (dotted-dashed line) to  second order in $T/E$.  }

\vskip1.0cm

The energy carried by the produced right-handed neutrinos or $\nu_R$ 
emissivity $q$ can be  obtained  from $\Gamma(E)$ according to the relation 
\be
   q =  \int {d^3 p \over \left( 2 \pi \right)^3}  \, E \, \Gamma(E)  \, .
   \label{emi}
\ee
The  $ \nu_R$ energy spectrum ($\propto  E^3 \, \Gamma(E)$) as a function of 
$E$ is shown in Fig.~5, where the exact numerical result, shown as solid line, 
is compared with the approximate analytical results. We observe an excellent 
agreement between the  exact and the  second order approximate solution.

Finally, we quote that in the degenerate limit $T \to 0$, $\Gamma(E)$ reduces 
to a very simple expression 
\be
   \Gamma(E) = \frac{\mu_\nu^2 \, m_\gamma^2    }{4 \pi  }  \, 
   \left(  \tilde{\mu}_\nu - E \right) \,  \theta( \tilde{\mu}_\nu - E) 
   \, ,
   \label{game8}
\ee
which is similar to the one obtained in Ref.~\cite{belma} for the damping
rate of a fermion in a QED dense relativistic plasma at zero temperature. 

\vskip1.0cm

\let\picnaturalsize=N
\def\picsize{5.0in}
\def\picfilename{figu5.eps}
\ifx\nopictures Y\else{\ifx\epsfloaded Y\else\input epsf \fi
\let\epsfloaded=Y
\centerline{\ifx\picnaturalsize N\epsfxsize \picsize\fi \epsfbox{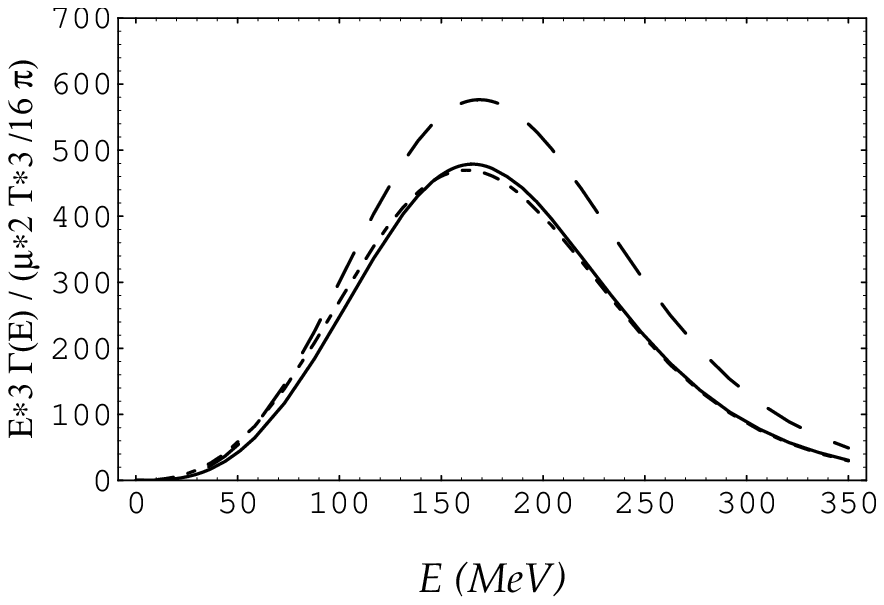}}}\fi

{\it Fig. 5   The energy spectrum  $\sim E^3 \Gamma(E)$ for right-handed 
         neutrinos  produced by the  spin flip transition 
         $\nu_L \to \nu_R$. The exact 
         (numerical ) result in the solid line is compared with the 
         approximate result (dashed line) to leading order in $T/E$ and 
         the approximate result (dotted  line) to  second order in $T/E$. }

\vskip1.0cm
\section{Plasmon to $\bar{\nu}_L \nu_R$ decay}\label{decay}

We now consider the contribution to the $\nu_R$ production rate arising from 
the plasmon decay process $\gamma \to \bar{\nu}_L \nu_R$. The rate is obtained 
from the general expression in Eq.~(\ref{game1}) by setting $E +  k_0< 0$ as 
corresponds to a negative energy $\nu_L$ or an outgoing $\bar{\nu}_L$. The 
angular integration over  the direction of  $\vec k$  leads now to the 
following kinematically restrictions 
\be 
   \omega &\equiv& - k_0 > 0
   \nonumber\\
   \omega^-  &\leq&  E \leq \omega^+  \, .
   \label{restric}
\ee
where  $\omega^{\pm} = (\omega \pm k)/2$.
These restrictions imply that for the plasmon decay process, the kinematically 
allowed region corresponds to the time-like region  $| \omega| > k$ and thus
the contribution from the photon spectral density in Eq.~(\ref{rolong}) arises 
solely from the poles. Using the identities
\be 
   1 + f(-\omega) &= & -f(\omega)
   \nonumber\\
   \tilde{n}_F\left(-E^\prime\right) &=& 1 - {1 \over e^{( E^\prime + 
   \tilde{\mu}_\nu)/T}+ 1 }  = 
   1 -   \tilde{n}_F (E^\prime)
   \, ,    \label{identi}
\ee
where $ \tilde{n}_F$ refers to the thermal distribution of $\bar{\nu}_L$, we 
can  obtain  the $\nu_R$ production rate from the plasmon decay process.  
Substituting the expressions for the fermion and photon propagators,
Eqs.~(\ref{self1}) and (\ref{fotopro}), into Eq.~(\ref{self2}) we obtain
\be 
   \Gamma(E) &=& \frac{\mu_\nu^2}{16\pi  E^2} \sum_{i= L,T}
   \int_0^{\infty} k dk 
   \,   \theta(\omega^+ - E)  \theta(E - \omega^-)  f(\omega_i(k)) 
   \left[ 1 -   \tilde{n}_F( \omega_i(k) - E) \right]  \nonumber\\
   & & C_i(E, - \omega_i(k)) Z_i(k) 
   \, , \label{game9}
\ee
where the sum is carried over the longitudinal and transverse modes. The  
$\nu_R$ emissivity can now be computed using Eq.~(\ref{emi}). In general, the 
$E$ and $k$ integrations have to be computed numerically, however, in many 
applications we can consider  that  the $\bar{\nu}_L$ neutrinos are absent from
the medium. Then, the Pauli blocking factor can be neglected, and the 
integration over the $\nu_R$ energy can be readily performed with the result 
\be 
   q  =    \frac{\mu_\nu^2}{96 \pi^3} 
   \int_0^{\infty} k^2 dk 
   \bigg[   & &{1 \over 2}  \omega_L(k)   
   \left( \omega_L^2(k) - k^2 \right)^2  f(\omega_L(k))  Z_L(k)\nonumber \\
   + & &\omega_T(k) \left( \omega_T^2(k) - k^2 \right)^2  
   f(\omega_T(k))  Z_T(k)
   \bigg]    \, . \label{qq}
\ee
The  remaining  integral still requires numerical computation. However,
analytic expressions can be  obtained in the limit cases $m_\gamma \ll T$ and 
$ m_\gamma   \gg T$. In order to discuss these limits, it is convenient to 
decompose the emissivity $q$ into its longitudinal and transverse parts 
$q_L$ and $ q_T$.  

The limit  $m_\gamma \ll T$  requires  ultrarelativistic temperatures and,  
according to Eq.~(\ref{mass}), an electron density bound by the condition
$e \tilde{\mu}_e \ll \sqrt{2} \pi T$. In this limit, the integral for the 
transverse emissivity in Eq.~(\ref{qq}) is dominated  by $k$ on the order of  
$T$. Thus, we can use the large $k$ limit of the dispersion relation in  
Eq. (\ref{reld2}) and the resulting integral can be evaluated analytically
with the result
\be 
   q_T \, = \,   \frac{\mu_\nu^2 \, \zeta(3)}{96 \pi^3}  \, m_\gamma^4 \, 
   T^3\, , 
\ee
where $\zeta(z)$ is the Riemann zeta function. 

For the longitudinal emissivity in the same limit $m_\gamma \ll T$, the 
contributions for large $k$ can be neglected.  The reason is that the 
substitution of the asymptotic expansion, Eq. (\ref{reld2}), in the factor 
$(\omega_L(k)^2 - k^2)^2$ produces a Gaussian cutoff in the integrals. Thus, 
the integral is dominated by $k$ on the  order of  $m_\gamma$ or smaller.  The 
Bose distribution can be approximated by $f(\omega_L) \sim T/\omega_L$. The 
integral  then involves only the scale $m_\gamma$ and  can be evaluated 
numerically 
\be 
   q_L \, = \,    0.28 \frac{\mu_\nu^2 }{96 \pi^3}  \, m_\gamma^6 \,  T
   \,  . 
\ee
We notice that in this limit the longitudinal emissivity is negligible as 
compared to the transverse emissivity, since it is suppressed by a factor
 $\left(m_\gamma /T \right)^2$.

The limit $T \ll m_\gamma$ is obtained  for an ultrarelativistic  electron 
density and low temperatures according to the relation $e \tilde{\mu}_e \gg  
\sqrt{2} \pi T$.  In this limit  the integrals in both  the longitudinal and  
transverse cases are dominated by  momenta small  compared to $m_\gamma$.  We 
can set $ k \to 0$ wherever possible except  that, according to the asymptotic  
expansion in  Eq. (\ref{reld1}),  the Bose distribution is approximated by 
$f  \to \exp[- (\omega_p  + \lambda k^2 /\omega_p)/T]$  with $\lambda = 3/5 , 
3/10$ for the transverse and longitudinal modes respectively.
The resulting integrals can be evaluated analytically with the results
\be 
   q_T \, &=& \,   \frac{\mu_\nu^2 }{768 \pi^3} 
   \left({125 \pi \over 27 } \right)^{1/2} \,  \omega_p^{11/2}  \,  T^{3/2} 
   e^{-\omega_p/T} \, , \nonumber \\
   q_L \, &=& \,   \frac{\mu_\nu^2 }{768 \pi^3} 
   \left({250 \pi \over 27 } \right)^{1/2} \,  \omega_p^{11/2}  \,  T^{3/2} 
   e^{-\omega_p/T} 
   \,  . 
\ee

\section{The $\nu_R$  emission in a  supernova }\label{super}

As a first application of our results we consider the  emission of right-handed
neutrinos immediately after a supernova core collapse.
The large mean free path of the right handed neutrinos compared to the core 
radius implies that the $\nu_R$'s would freely fly away from the supernova.
Therefore, the core luminosity for $\nu_R$ emission can be simply  computed as
\be
   Q_{\nu_R} \, = \,  V \, q 
   \label{lumi}  \, ,
\ee
where $V$ is the plasma volume and $q$ is the $\nu_R$ emissivity computed from 
Eq. (\ref{emi}). To make a numerical estimate, we shall adopt a simplified  
picture of the inner core, corresponding to the the average parameters of 
SN1987A~\cite{mohapatra,burro}. Consequently, we take a constant density 
$ \rho \approx 8 \times 10 ^{14} \, {\rm g/cm}^3$, a volume
$V \approx  8 \times 10^{18} \,  {\rm cm}^3$, an electron to baryon
ratio $Y_e \simeq Y_p \simeq 0.3$, and   temperatures in the range
$T = 30 \sim 60 \, {\rm MeV}$. This corresponds to a degenerate electron gas 
with a chemical potential $\tilde{\mu}_e$ ranging from $307$ to $280 \,  
{\rm MeV}$. For the left-handed neutrino we take $\tilde{\mu}_\nu \approx 160 
\, {\rm MeV}$. Using this values in  Eqs.~(\ref{game2}) and (\ref{lumi}), we 
obtain by numerical integration
\be
   Q_{\nu_R} = \left( { \mu_\nu \over \mu_B  }  \right)^2  \,(0.7-4.3)\times
   10^{76}\,
   {\mbox e}{\mbox r}{\mbox g}{\mbox s}/{\mbox s}{\mbox e}{\mbox c} \,  ,
   \label{numlum}
\ee
for $T$ ranging from $30$ to $60$ MeV. The main contribution to this result 
arises from the $\nu_L \to \nu_R$  flip process,  whereas the  contribution 
from the  $\gamma \to \bar{\nu}_L \nu_R$ decay is smaller by  two orders of 
magnitude. This  result is in agreement with the observation first made by 
Fukugita and Yazaki~\cite{yazaki} who noticed that for cosmological and 
astrophysical scenarios the plasmon decay process is subdominant as compared 
to the chirality flip process. Moreover, an estimate of the $\nu_R$ luminosity
derived from the approximate solutions discussed in section~\ref{flip} are 
surprisingly accurate, they differ from the result in Eq. (\ref{numlum}) by 
less than $2\%$.  

Assuming that the emission of $\nu_R$'s lasts approximately for $1$ sec., the 
luminosity bound is $Q_{\nu_R}\leq 10^{53}$ ergs/sec. which places the upper 
limit on the neutrino magnetic moment
\be
   \mu_\nu < (0.1-0.4) \times 10^{-11}\mu_B\, .
   \label{momm}
\ee
This upper bound slightly improves the result previously obtained by Barbieri 
and Mohapatra~\cite{barbi}. As mentioned before, these authors consider
the helicity flip scattering $\nu_L e \to \nu_R  e$  to order $e^4$ 
introducing the Debye mass in the photon propagator as an infrared regulator. 

A word of caution should be mentioned in relation to  the   result  in
Eq.~(\ref{momm}). It has been pointed out by Voloshin~\cite{volo2} that the
$\nu_R$'s produced by the magnetic moment interaction could undergo resonant
conversion back into $\nu_L$'s through spin rotation  in the magnetic field of
the supernova core, with the subsequent trapping of the $\nu_L$'s  by the 
external layers. If this is the case, then the  bound  in Eq.~(\ref{momm}) 
becomes meaningless. However, the core density is rather high and the matter 
effect might dominate over the $\mu_\nu\,B$ term, suppressing the flip back of 
$\nu_R$ to $\nu_L$~\cite{mohapatra}.

Recently, another mechanism for the neutrino chirality flip has been proposed,
which occurs via the \v{C}erenkov emission or absorption of plasmons in
the supernova core~\cite{mohanty}. Since the photon dispersion relation in 
a relativistic plasma shows a space-like branch for the longitudinal mode,
the \v{C}erenkov radiation of the plasmon is, in principle, kinematically 
allowed~\cite{jc}. However, this mode develops a
large imaginary  part, which implies that the Landau damping mechanism
acts to preclude its propagation as we have discussed. Consequently, we 
think that no better than the quoted limit in Eq.~(\ref{momm}) can be derived 
by this type of neutrino chirality flip processes in a supernova core.

\section{Early Universe}\label{early}

As a second application of our results, we now consider the production of
right-handed neutrinos during the evolution of the early universe. If
the rate of production of these neutrinos is able to maintain them
in thermal contact with the rest of the particles in the plasma, then
they will contribute to the effective number of degrees of freedom
until their final decoupling. This could in principle affect primordial
nucleosynthesis. In order to prevent right-handed neutrinos from being in
thermal equilibrium, we need to require that their average production rate be 
less than the Hubble rate at all times until the neutrino freeze-out epoch.
During the radiation dominated era, the Hubble rate was 
\be
   H=\frac{T^2}{m_{Plank}}\left(\frac{4\pi^3}{45}g_*\right)^{1/2}\, ,
   \label{Hubb}
\ee
with $g_*\simeq 10.75$ the effective number of degrees of freedom at the
nucleosynthesis epoch. On the other hand, the average right-handed neutrino
production rate can be obtained from Eq.~(\ref{game2}) (neglecting the
contribution from the plasmon decay process) by averaging with
an equilibrium distribution 
\be
   n_{\nu}(E)=\frac{1}{\exp(E/T)+1}\, ,
   \label{eqdn}
\ee
appropriate for the early universe where the chemical potentials should be
negligible. Therefore, the condition to avoid populating the right-handed
neutrino component becomes
\be
   \langle \Gamma \rangle = 5.78\times 10^{-4} T^3\mu_{\nu}^2 <
   \frac{T^2}{m_{Plank}}\left(\frac{4\pi^3}{45}g_*\right)^{1/2}\, ,
\ee
where for nucleosynthesis, $100$ MeV $> T >1$ MeV. The most stringent bound on 
$\mu_{\nu}$ is obtained for the highest possible temperature and thus we
take $T=100$ MeV which yields
\be
   \mu_\nu < 2.9\times 10^{-10}\mu_B\, .
\ee
This result has to be compared to that from Ref.~\cite{elmfors} where the
use of a {\it full one-loop} approximation to the photon polarization functions
is used, instead of Eqs.~(\ref{prop2}). This choice leads to an upper bound one 
order of magnitude smaller than the above. One should notice however that
when a plasma is such that the largest energy scale available is set by the
temperature (or density), as in the present scenario, it is necessary to select
the leading temperature (density) contributions out of perturbative 
calculations and sum these up into effective vertices and propagators in order 
to extract from them meaningful quantities. In this manner, one can ensure 
that the leading perturbative corrections are effectively taken into 
account~\cite{brat1}. Therefore, we conclude that the bound on the neutrino 
magnetic moment set by nucleosynthesis constraints is not nearly as stringent 
as that set by the analysis of a supernova core collapse.

\section{Conclusions}\label{conclu}

In conclusion, we have shown that the scattering processes mediated by 
effective plasma photons allows for the efficient conversion of $\nu_L$ into 
$\nu_R$. In this work, plasma effects are consistently taken into account by
means of the resummation method of Braaten and Pisarski within the 
real-time formulation of TFT. For soft values of
the energy, the production rate of $\nu_R$ differs significantly  from that 
obtained by a constant Debye mass screening prescription.  However,
corrections to the integrated luminosity are small and for this reason, our 
upper bound on the neutrino magnetic moment does not differ significantly 
from the one obtained by Barbieri and Mohapatra~\cite{barbi}. Knowledge of an 
accurate expression for the $\nu_R$ production rate, as given in 
Eq.~(\ref{game2}), could be of importance in a detailed analysis of supernova
processes. We also obtain another constraint on $\mu_{\nu}$ by considering the
possible effect of $\nu_R$ production in the early universe. The upper bound 
imposed by the analysis of SN1987A is two orders of magnitude smaller than 
the one obtained from the nucleosynthesis constraint.

\section{Appendix A}

In this appendix, we want to explicitly show that the transverse photon 
contribution, $\Gamma_T(E)$, to the right-handed neutrino production rate, 
$\Gamma(E)$, is free from infrared divergences. To this end, we refer back to 
Eqs.~(\ref{bet2}) and (\ref{game2}) that give the explicit expression for 
$\Gamma_T(E)$. Let us write $x=\omega /k$ thus, when $x \to 0$, we can write
\be
   1+f(kx)\to\frac{T}{kx}\, ,
   \label{ap1}
\ee
where $f(z)$ is the photon statistical distribution. Notice that in this limit,
$\beta_T(k,x)[1+f(kx)]$ remains finite, unless we also take $k\to 0$. Let us 
write
\be
   \beta_T(k,x)=\frac{1}{k^2(1-x^2)}\tilde{\beta}_T(k,x)\, ,
   \label{ap2}
\ee
with the definition
\be
   \tilde{\beta}_T(k,x)\equiv\frac{k^2m_{\gamma}^2x(1-x^2)^2/2}
   {\left[ k^2(1-x^2) + m_{\gamma}^2(x^2+\frac{x}{2}(1-x^2)\ln\left(
   \frac{1+x}{1-x}\right) )\right]^2 + \left[\pi m_\gamma^2 x\frac{(1-x^2)}
   {2}\right]^2}\, .
   \label{ap3}
\ee
Notice that for $x\ll 1$,
\be
   \frac{\tilde{\beta}_T(k,x)}{x}\to \frac{m_\gamma^2}{2}\frac{k^2}
   {k^4+(\pi m_\gamma^2 x/2)^2}\, .
   \label{app4}
\ee
Therefore, in the limit $k\to 0$,
\be
   \frac{\tilde{\beta}_T(k,x)}{x}\to \delta(x)\, .
   \label{app5}
\ee
In this manner, we can write the contribution to $\Gamma_T(E)$ from the soft 
($k\to 0, \omega\to 0$) region as
\be
   \Gamma_T^{soft}(E)&\approx&\frac{\mu_{\nu}^2 T}{16\pi E^2}n_F(E)
   \int_0^{m_\gamma}\!\!dk\, k\,\theta(2E-k)
   [(2E)^2-k^2]\nonumber\\
   &=&\frac{\mu_{\nu}^2 T}{16\pi E^2}n_F(E)
   \left\{ \begin{array}{ll}
           m_\gamma^2(2E^2-m_\gamma^2/4), & 2E\geq m_\gamma \\
           4E^4,                          & 2E < m_\gamma
           \end{array}
   \right.\, ,
   \label{app6}
\ee
where in the integration we have set as the upper limit the soft scale 
$m_\gamma$. This choice is somewhat arbitrary but it does not matter as long
as for the hard contribution to $\Gamma_T(E)$ we integrate $k$ from
$m_\gamma$. Eq.~(\ref{app6}) is thus explicitly infrared finite.

\section{Appendix B}

The second order correction  in ($T/E$)  to $\Gamma(E)$ for the  quirality 
flip $\nu_L \to \nu_R$ reaction can also be calculated analytically.  As 
discussed in section~\ref{flip}, the leading order term is obtained by 
setting  $\omega , k \to 0$ as compared with $E$ wherever possible, except for
the Bose and Fermi distributions. In  the soft-momentum  region, the condition
$\omega , k \ll E$  is valid everywhere, and consequently 
there are no $T/E$ corrections to the result in Eq.~(\ref{game4}). 

The hard-momentum region  $(k > q^* )$  was  decomposed  into  
a  (I)  low $(\omega <  q^* )$  and  (II) high frequency $(\omega > q^* )$  
regions. In the low  frequency region we have $\omega \ll E$ and only  $k/E$ 
corrections  have to  be computed. It is a simply exercise to show that the 
only corrections are of order  $\left(T/E\right)^2$, hence the result in  
Eq.~(\ref{game5}) is not modified in the next order. Finally, in the  high 
frequency region we compute corrections of order $k/E$ and  $\omega/E$ to the 
result in Eq.~(\ref{game6}). In  particular, instead of Eq.~(\ref{traza2}), 
we approximate the functions in Eq.~(\ref{trace}) by
\be 
   C_T(E,K)\approx - C_L(E,K)\approx {4K^4\over k^2 } 
   \left( E^2 + E \omega \right)
   \, . 
\ee

The correction to the result in  Eq.~(\ref{game6}) is given in terms of 
PolyLog functions as 
\be
\Gamma(E) &=&\frac{\mu_\nu^2 \, m_\gamma^2  T  }{2 \pi  }  n_F(E )
\bigg[  \left( {T \over 8 E} \right) \left( - \pi^2 + 7  Li_2\left(e^{-E/T}\right) +
7 Li_2\left(- e^{\tilde{\mu}_\nu/T}\right) - 8   Li_2\left(- e^{(\tilde{\mu}_\nu - E )/T}\right)
 \right)  \nonumber \\
&+&  7 \left({E \over T} \right)  \ln \left( { 1  +  e^{\tilde{\mu}_\nu/T }  \over 
1 -  e^{- E/T } }\right)  - {7 \over 2}   \left({E \over T} \right)^2 
\bigg]
\, . 
\ee
This corrections improve the  approximate analytical solution given in 
Eq.~(\ref{game7})
when compared with the exact result in Eq.~(\ref{game2})  for energies 
$E {\ \lower-1.2pt\vbox{\hbox{\rlap{$>$}\lower5pt\vbox{\hbox{$\sim$}}}}\ } 
T, \tilde{\mu}_\nu$, as shown in Figs. 4 and 5.

\section*{Acknowledgments}

This work was  supported in part by Universidad Nacional 
Aut\'onoma de M\'exico under Grants DGAPA-IN117198 and DGAPA-IN10389, 
and by CONACyT-M\'exico under Grants 3097 p-E and I27212-E.

\end{document}